\documentclass[12pt]{article}

\usepackage{latexsym,amsmath,amscd,amssymb,framed,graphics,mathrsfs}
\usepackage{enumerate}

\usepackage{graphicx}

\usepackage[square,authoryear]{natbib}
\usepackage{url}
\usepackage{colordvi}

\usepackage[all]{xy}

\newcommand{\bfi}{\bfseries\itshape}

\makeatletter

\@addtoreset{figure}{section}
\def\thefigure{\thesection.\@arabic\c@figure}
\def\fps@figure{h, t}
\@addtoreset{table}{bsection}
\def\thetable{\thesection.\@arabic\c@table}
\def\fps@table{h, t}
\@addtoreset{equation}{section}

\makeatother

\begin{document}

\newtheorem{theorem}{Theorem}[section]
\newtheorem{definition}[theorem]{Definition}
\newtheorem{lemma}[theorem]{Lemma}
\newtheorem{remark}[theorem]{Remark}
\newtheorem{proposition}[theorem]{Proposition}
\newtheorem{corollary}[theorem]{Corollary}
\newtheorem{example}[theorem]{Example}

\def\below#1#2{\mathrel{\mathop{#1}\limits_{#2}}}



\title{Higher order Lagrange-Poincar\'e and Hamilton-Poincar\'e reductions}
\author{Fran\c{c}ois Gay-Balmaz$^{1}$, Darryl D. Holm$^{2}$, and Tudor S. Ratiu$^{3}$}
\addtocounter{footnote}{1} 
\footnotetext{Laboratoire de 
M\'et\'eorologie Dynamique, \'Ecole Normale Sup\'erieure/CNRS, Paris, France. 
\texttt{gaybalma@lmd.ens.fr}
\addtocounter{footnote}{1} }
\footnotetext{Department of Mathematics, Imperial College, London SW7 2AZ, UK.
Partially supported by Royal Society of London, Wolfson Award. 
\texttt{d.holm@ic.ac.uk}
\addtocounter{footnote}{1} }
\footnotetext{Section de
Math\'ematiques and Bernoulli Center, \'Ecole Polytechnique F\'ed\'erale de
Lausanne. CH--1015 Lausanne. Switzerland. Partially supported by Swiss NSF grant 200020-126630. 
\texttt{tudor.ratiu@epfl.ch}
\addtocounter{footnote}{1} }


\date{ }
\maketitle

\makeatother
\maketitle


\begin{center}\large J. Braz. Math. Soc. \textbf{42}(4), (2011), 579--606
\end{center}

\begin{abstract}
Motivated by the problem of longitudinal data assimilation, e.g., in the registration of a sequence of images, we develop the higher-order framework for Lagrangian and Hamiltonian reduction by symmetry in geometric mechanics. In particular, we obtain the reduced variational
principles and the associated Poisson brackets. The
special case of higher order Euler-Poincar\'e and
Lie-Poisson reduction is also studied in detail.
\end{abstract}
\medskip


\tableofcontents


\section{Introduction}\label{introduction}

\paragraph{Background.}
Many interesting mechanical systems, such as the incompressible fluid, the rigid body, the KdV equation, or the Camassa-Holm equations can be written as the \textit{Euler-Poincar\'e equations} on a Lie algebra $\mathfrak{g}$ of a Lie group $G$. The corresponding Hamiltonian formulations are given by \textit{Lie-Poisson equations} obtained by Poisson reduction of the canonical Hamilton equations on $T^*G$.

A more general situation occurs if the original configuration space is not a Lie group, but a configuration manifold $Q$ on which a Lie group $G$ acts freely and properly, so that $Q\rightarrow  Q/G$ becomes a principal $G$-bundle. Starting with a Lagrangian system on $TQ$ invariant under the tangent lifted action of $G$, the reduced equations on $(T Q)/G$, appropriately identified, are the \textit{Lagrange-Poincar\'e equations} derived in \cite{CeMaRa2001}. 
Similarly, if we start with a Hamiltonian system on $T^*Q$, invariant under the cotangent lifted action of $G$, the resulting reduced equations on $(T^*Q)/G$ are called the \textit{Hamilton-Poincar\'e equations}, \cite{CeMaPeRa2003}, with an interesting Poisson bracket, the \textit{gauged Lie-Poisson} structure, involving a canonical bracket, a Lie-Poisson bracket, and a curvature term.

\paragraph{Goals.}
The goal of this paper is to present the extension of this picture to the higher order case, that is, the case when the Lagrangian function is defined on the $k^{th}$-order tangent bundle $T^{(k)}Q$ and thus depends on the first $k^{th}$-order time derivatives of the curve. We thus derive the \textit{$k^{th}$-order Lagrange-Poincar\'e equations} on $T^{(k)}Q/G$ and obtain the \textit{$k^{th}$-order Euler-Poincar\'e equations} on $T^{(k)}G/G\simeq k\mathfrak{g}$ in the particular case $Q=G$, together with the associated constrained variational formulations.

On the Hamiltonian side, using the Legendre transform $T^{(2k-1)}Q\rightarrow T^*\left(T^{(k-1)}Q\right)$ associated to the Ostrogradsky momenta, we obtain what we call the \textit{Ostrogradsky-Hamilton-Poincar\'e equations} on $T^*\left(T^{(k-1)}Q\right)/G$ and, in the particular case $Q=G$, the \textit{Ostrogradsky-Lie-Poisson equations}.

\paragraph{Motivation and approach.}
Our motivation for making these extensions to higher order of the fundamental representations of dynamics in geometric mechanics is to cast light on the options available in this framework for potential applications, for example, in longitudinal data assimilation. However, these applications will not be pursued here and we shall stay in the context of the initial value problem, rather than formulating the boundary value problems needed for the applications of the optimal control methods, say, in longitudinal data assimilation.  For further discussion of the motivation for developing the higher-order framework for geometric mechanics in the context of optimal control problems, in particular for registration of a sequence of images, see \cite{GBHoMeRaVi2011}.

\section{Geometric setting}\label{geomset-sec}

We shall begin by reviewing the definition of higher order tangent bundles $T^{(k)}Q$, the connection-like structures defined on them, and the description of the quotient of $T^{(k)}Q$ by a free and proper group action. For more details and explanation of the geometric setting for higher order variational principles in the context that we follow here, see \cite{CeMaRa2001}. We also recall the formulation of the $k^{th}$-order Euler-Lagrange equations and the associated Hamiltonian formulation obtained through the Ostrogradsky momenta. We refer to \cite{dLRo1985} for the geometric formulation of higher order Lagrangian and Hamiltonian dynamics.

\subsection{Higher order tangent bundles}\label{kth_order_tgt_bundles}

The \textit{${k^{th}}$-order tangent bundle} $\tau^{(k)}_Q : T^{(k)}Q
\rightarrow Q$ of a manifold $Q$ is defined as the set of equivalence classes of
$C^k$ curves in $Q $ under the equivalence relation that
identifies two given curves $q_i(t), i = 1,
2$, if $q_1(0) = q_2(0) = q_0$ and in any local chart we have 
$q^{(l)}_1(0) =q^{(l)}_2(0)$, for $l = 1, 2,\ldots , k$, where $ q^{(l)}$  denotes the
derivative of order $l$. The equivalence class of the curve $q(t)$ at $q_0 \in Q$ is denoted $[q]_{q_0}^{(k)}$. The projection
\[ 
\tau^{(k)}_Q : T^{(k)}Q \rightarrow Q \quad  \mbox{is given by} \quad
\tau^{(k)}_Q\left([q]_{q_0}^{(k)}\right) = q_0.
\]

It is clear that $T^{(0)}Q = Q$, $T^{(1)}Q = TQ$, and that, for $0 \leq l < k$, there
is a well defined fiber bundle structure
\[ 
\tau^{(k,l)}_Q : T^{(k)}Q \rightarrow T^{(l)}Q, \quad  \mbox{given by} \quad
\tau^{(k,l)}_Q\left([q]_{q_0}^{(k)}\right) = [q]_{q_0}^{(l)}.
\]
Apart from the cases $k = 0$ and $k=1$, the bundles  $T^{(k)}Q$ are not vector bundles. We shall use the natural coordinates $(q,\dot q,..., q^{(k)})$ on $T^{(k)}Q$ induced by a coordinate system on $Q$.

A smooth map $f : M \rightarrow N$ induces a map
\begin{equation}
\label{kth_order_tangent_map} 
T^{(k)}f : T^{(k)}M \rightarrow T^{(k)}N \quad  \mbox{given by} \quad
T^{(k)}f\left([q]_{q_0}^{(k)}\right) := [f\circ q]^{(k)}_{f(q_0)}.
\end{equation} 
In particular, a group action $\Phi : G \times Q \rightarrow Q$ 
naturally lifts to a group action
\begin{equation}
\label{kth_order_action}
\Phi^{(k)} : G \times T^{(k)}Q \rightarrow T^{(k)}Q \quad  \mbox{given by}
\quad \Phi^{(k)}_g\left([q]_{q_0}^{(k)}\right):=T^{(k)}\Phi_g \left( [q]_{q_0}^{(k)} \right) = \left[\Phi_g \circ q\right]^ {(k)}_{\Phi_g(q_0) } .
\end{equation} 
When the action $\Phi$ is free and proper we get a principal $G$-bundle   $T^{(k)}Q\rightarrow \left( T^{(k)}Q\right) /G$. The quotient
$\left( T^{(k)}Q\right) /G$ is a fiber bundle over the base $Q/G$. The class of the element $[q]_{q_0}^{(k)}
\in T^{(k)}_{q_0}Q$ in the
quotient $\left( T^{(k)}Q\right) /G$ is denoted $\left[[q]_{q_0}^{(k)}\right]_G$.

\subsection{Higher order Euler-Lagrange equations}\label{sec 2.2}

Consider a Lagrangian $L:T^{ (k) }Q\rightarrow \mathbb{R}$ defined on the $k^{th}$-order tangent bundle. We will often use the local notation $L\left( q, \dot q,....,  q^{ (k) }\right)$ instead of the intrinsic one $L\left(\left[q\right]^{(k)}_{q_0}\right)$. A curve
$q: [t_0, t_1] \rightarrow Q$
is a critical curve of the action
\begin{equation}\label{Euler-Lagrange_action}
 \mathcal{J}[q] = \int_{t_0}^{t _1} L \left( q(t), \dot q(t),....,  q^{ (k) }(t) \right)  dt  
\end{equation}
among all curves  $q(t) \in Q$ whose first $(k-1)$ derivatives $q^{(j)}(t_i)$, $i=0,1$, $j=0,..., k-1$, are fixed at the endpoints 
if and only if $q(t)$ is a solution of the \textit{$k^{th}$-order Euler-Lagrange equations}
\begin{equation}\label{EL-eqns}
\sum_{j=0}^k (-1)^j \frac{d^j}{dt^j}  \frac{\partial  L}{\partial  q^{ (j) }}=0.
\end{equation}

These equations follow from Hamilton's variational principle,
\[
\delta \int_{t _0 }^{ t _1} L\left( q(t), \dot q(t),....,  q^{ (k) }(t) \right)  dt  =0
.
\]
In the $\delta$-notation, an infinitesimal variation of the curve $q(t)$ is denoted by $\delta q(t) $ and defined by the \emph{variational derivative}, 
\begin{equation}
\label{var-deriv-def}
\delta q(t) : 
=\left.\frac{d}{d \varepsilon}\right|_{\varepsilon =0 } 
q(t, \varepsilon),
\end{equation}
where $q(t,0)=q(t)$ for all $t$ for which the curve is 
defined and $\frac{\partial^j q}{\partial t^j}(t_i,\varepsilon) = q^{(j)}(t_i)$, for
all $\varepsilon$, $j=0,1,\ldots, k-1$, and $i=0,1$. Thus
$\delta q ^{ (j) }(t_0)=0=\delta q ^{ (j) }(t_1)$ for all $j=0,...,k-1$.

\paragraph{Examples: Riemannian cubic polynomials and generalizations.}
As originally introduced in \cite{NoHePa1989}, \textit{Riemannian cubic
polynomials} (or \textit{$2$-splines}) generalize Euclidean splines to Riemannian manifolds.
Let $(Q,\gamma )$ be a Riemannian manifold and $\frac{D}{Dt}$ be the
covariant derivative along curves associated with the Levi-Civita
connection $\nabla$ for the metric $\gamma$.
The Riemannian cubic polynomials  are defined as minimizers of the
functional $\mathcal{J}$ in \eqref{Euler-Lagrange_action} for the Lagrangian $L: T^{(2)}Q\rightarrow \mathbb{R}$ defined by
\begin{equation}
L(q,\dot{q},\ddot{q}) := \frac{1}{2}\gamma_q\left(\frac{D}{Dt}\dot
q,\frac{D}{Dt}\dot q\right).
\end{equation}
This Lagrangian is well-defined on the second-order tangent bundle since, in coordinates,
\begin{equation}
\frac{D}{Dt}\dot{q}^k=
\ddot{q}^k + \Gamma_{ij}^k(q) \dot{q}^i\dot{q}^j ,
\end{equation}
where $\Gamma_{ij}^k(q)$ are the Christoffel symbols at the point
$q$ of the metric $\gamma$ in the given basis.
These Riemannian cubic polynomials have been generalized to the
so-called \textit{elastic $2$-splines} through the following class of Lagrangians
\begin{equation}\label{2_splines_Lagr}
L(q,\dot{q},\ddot{q}) : = \frac{1}{2} \gamma_{q} \left(
\frac{D}{Dt}\dot q, \frac{D}{Dt}\dot q\right)  + \frac{\tau^2}{2}  \gamma_{q}
(\dot{q},\dot{q}) ,
\end{equation}
where $\tau$ is a real constant, see \cite{HuBl2004}.
Another extension are the \textit{$k^{th}$-order Riemannian splines}, or \textit{geometric $k$-splines}, where
\begin{equation}\label{k_splines_Lagr}
L\left( q,\dot q,..., q^{(k)} \right) :=  \frac{1}{2} \gamma_{q}
\left( \frac{D^{k-1}}{Dt^{k-1}}\dot q,\frac{D^{k-1}}{Dt^{k-1}}\dot q
\right) ,
\end{equation}
for $k >2$.
As for the Riemannian cubic splines, $L$ is well-defined on $T^{(k)}Q$.
Denoting by $R$ the curvature tensor defined as
$R(X,Y)Z= \nabla _X \nabla _Y Z- \nabla _Y \nabla _X Z- \nabla
_{[X,Y]}Z$, the Euler-Lagrange equation for elastic $2$-splines ($k=2$) reads
\begin{equation} \label{ELeqns-T2}
\frac{D^3}{Dt^3}\dot q(t)+ R \left( \frac{D}{Dt}\dot q(t),\dot q(t)
\right) \dot q(t)= \tau ^2 \frac{D}{Dt}\dot q (t) ,
\end{equation}
as proven in \cite{NoHePa1989}, \cite{HuBl2004}. For the higher-order Lagrangians \eqref{k_splines_Lagr},
the Euler-Lagrange equations read, \cite{CaSLCr1995},
\begin{equation}
\frac{D^{2k-1}}{Dt^{2k-1}}\dot q(t) + \sum_{j=2}^k (-1)^j R
\left(\frac{D^{2k-j-1}}{Dt^{2k-j-1}}\dot q(t),
\frac{D^{j-2}}{Dt^{j-2}}\dot q(t)\right)\dot q(t) =0.
\end{equation}

\begin{remark}\label{important_remark}{\rm It is important to note that the higher order Lagrangians \eqref{2_splines_Lagr}, \eqref{k_splines_Lagr} are functions defined on the manifolds $T^{(2)}Q$ and $T^{(k)}Q$, respectively, and not on curves $q(t)\in Q$. Therefore, the notation $ \frac{D}{Dt}\dot q$ in formula $\eqref{2_splines_Lagr}$ means the expression in terms of $\dot q$ and $\ddot q$ seen as \emph{independent elements} in the manifold $T^{(2)}Q$.}
\end{remark}

\subsection{Quotient space and reduced Lagrangian}

We now review the geometry of the quotient space $\left( T^{(k)}Q\right) /G$ relative to the lifted action $\Phi^{(k)}$, in preparation for the reduction processes we shall present in the next sections.

\paragraph{The quotient space $\left(T^{(k)}Q\right)/G$.} Consider a free and proper right (resp. left) Lie group action $\Phi$ of $G$ on $Q$. Let us fix a principal connection $\mathcal{A}$ on the principal bundle $\pi:Q\rightarrow Q/G$, that is, a one-form $\mathcal{A}\in\Omega^1(Q,\mathfrak{g})$ such that
\[
\mathcal{A}(\xi_Q(q))=\xi,\quad \Phi_g^*\mathcal{A}=\operatorname{Ad}_{g^{-1}}\circ\mathcal{A},\quad\text{resp.}\quad \Phi_g^*\mathcal{A}=\operatorname{Ad}_g\circ\mathcal{A},
\]
where $\xi_Q$ is the infinitesimal generator associated to the Lie algebra element $\xi$.
Recall that by choosing a principal connection $ \mathcal{A} $ on the principal bundle $\pi: Q\rightarrow Q/G$, we can construct a vector bundle isomorphism
\begin{equation}\label{alpha_1}
\alpha_ \mathcal{A} : 
(TQ)/G\rightarrow T(Q/G)\oplus \operatorname{Ad}Q,\quad \alpha_ \mathcal{A} \left( \left[[q]_{q _0}\right] _G  \right) 
:= \left( T\pi \left( [q]_{q _0} \right) , \left[ q _0 , \mathcal{A} \left( [q]_{q _0}\right) \right] _G \right),
\end{equation}
where $\oplus$ denotes the Whitney sum and the adjoint bundle $ \operatorname{Ad}Q \rightarrow Q/G$ is the vector bundle defined by the quotient space $ \operatorname{Ad}Q:= \left( Q\times \mathfrak{g}  \right)/G$ relative to the diagonal action of $G$.

We now recall from \cite{CeMaRa2001}  how this construction generalizes to the case of ${k^{th}}$-order tangent bundles. The covariant derivative of a curve $\sigma (t)=[q(t), \xi (t)] _G \in \operatorname{Ad}Q$ relative to a given principal connection $\mathcal{A}$ is given by
\begin{equation}
\label{cov_der_general}
\frac{D}{Dt}  \sigma (t)=\frac{D}{Dt} [q(t), \xi (t)] _G= \left[q(t),\dot{\xi} (t)\pm [\mathcal{A} ( \dot q(t)), \xi (t) ]\right] _G,
\end{equation}
where the upper (resp. lower) sign in $(\pm)$ corresponds to a right (resp. left) action. In the particular case when $\sigma (t)=[q(t),\mathcal{A} ( \dot q(t))]_G $, we have
\begin{align}
\frac{D}{Dt} \sigma (t)&= \left[q(t),\dot{\xi} (t) \right] _G =: \left[q(t),\xi _2 (t) \right] _G,\label{eq_1}\\
\frac{D^2}{Dt^2 }\sigma (t)&=\left[q(t),\ddot{\xi} (t)\pm [\xi (t), \dot\xi (t) ]\right] _G =:\left[q(t),\xi _3 (t) \right] _G,
\end{align}
and, more generally,
\begin{equation}\label{xi_dot}
\frac{D^l}{Dt^l} \sigma (t)=\left[q(t),\xi_{l+1} (t)  \right] _G,\quad \xi_{l+1} (t) = \dot{\xi_l}(t) \pm [\xi(t) , \xi _l(t) ],
\end{equation}
where $\xi_1(t):=\xi(t)$ and $l=1, \ldots, k$.
The bundle isomorphism  $\alpha _ \mathcal{A}^{(k)} : \left( T^{ (k) }Q \right) /G \rightarrow T^{ (k) } (Q/G) \oplus k \operatorname{Ad}Q$ that generalizes \eqref{alpha_1} to the $k^{th}$-order case is defined by
\begin{equation}\label{quotmapLP}
\alpha _ \mathcal{A} ^ {(k) } \left( \left[[ q]^{ (k) }_{ q _0 }\right] _G \right)= \left( T^{ (k) } \pi \left(  [ q]^{ (k) }_{ q _0 }\right),  \sigma (0), \left.\frac{D}{Dt}\right|_{t=0} \!\! \sigma (t),\left.\frac{D^2}{Dt^2}\right|_{t=0} \!\!\sigma  (t),..., \left.\frac{D^{k-1}}{Dt^{k-1}}\right|_{t=0}  \!\!\sigma (t)\right),
\end{equation}
where $\sigma (t):=\left[q(t), \mathcal{A} ( \dot q(t)) \right] _G $, $q(t)$ is any curve representing 
$[q]^{(k)}_{q _0}\in T^{(k)}Q$ such that $q(0)=q _0$, and $k\operatorname{Ad}Q$ denotes the Witney sum of $k$ copies of the adjoint bundle. We refer to \cite{CeMaRa2001} for further information and proofs. We will use the suggestive notation
\begin{equation}\label{shortland_notation}
\left( \rho , \dot \rho , ..., \rho ^{ (k)}, \sigma ,\dot\sigma , ...,\sigma ^{(k-1)} \right) =\alpha _ \mathcal{A} ^ {(k) } \left( \left[[ q]^{ (k) }_{ q _0 }\right] _G \right)
\end{equation}
for the reduced variables.

\begin{remark}\label{important_remark}{\rm It is important to observe that the notation $\left.\frac{D^j}{Dt^j}\right|_{t=0} \sigma  (t)$ in the quotient map \eqref{quotmapLP} stands for the intrinsic expression obtained via
\eqref{eq_1}--\eqref{xi_dot} from the element $\left[[ q]^{ (k) }_{ q _0 }\right] _G \in \left(T^{(k)}Q\right)/G$. 

In \eqref{shortland_notation} the dot notations on $\rho$ and $\sigma$ have not the same meaning: $\rho , \dot \rho , ..., \rho ^{ (k)}$ are natural coordinates on $T^{(k)}(Q/G)$, whereas $\dot\sigma , ...,\sigma ^{(k-1)} $ are elements in $\operatorname{Ad}Q$, all seen as independent variables. When dealing with curves, $\rho^{(i)}$ really means the \textit{ordinary time derivative in the local chart}, whereas $\sigma^{(i)}$ means the \textit{covariant derivative} $\frac{D^i}{Dt^i}\sigma$.}
\end{remark}

\paragraph{The reduced Lagrangian.} If $L:T^{ (k) }Q \rightarrow \mathbb{R}  $ is a $G$-invariant $k^{th}$-order Lagrangian, then it induces a Lagrangian $\ell$ defined on the quotient space $\left( T^{ (k) }Q \right) /G$. If a connection is chosen, then we can write the reduced Lagrangian as
\[
\ell=\ell \left( \rho , \dot \rho , ..., \rho ^{ (k)}, \sigma , \dot\sigma , ..., \sigma ^{ (k-1)} \right): T^{ (k) }(Q/G)\oplus k \operatorname{Ad}Q\rightarrow \mathbb{R} .
\]

\paragraph{The case of a Lie group.} Let us particularize the map $\alpha _ \mathcal{A} ^{(k)}$ to the case where $Q$ is the Lie group $G$. The adjoint bundle $\operatorname{Ad}G$ can be identified with the Lie algebra $\mathfrak{g}$ via the isomorphism
\begin{equation}\label{Adjoint_bundle_identification}
\operatorname{Ad}G\rightarrow \mathfrak{g}  ,\quad [ g, \xi ]_G\mapsto \eta :=\operatorname{Ad}_g \xi ,\quad\text{resp.} \quad [ g, \xi ]_G\mapsto \eta :=\operatorname{Ad}_{g^{-1} } \xi.
\end{equation} 
The principal connection is the Maurer-Cartan connection
\[
\mathcal{A} ( v_g)= g ^{-1} v_g,\quad\text{resp.} \quad \mathcal{A} ( v_g)= v_g g ^{-1}
\]
and one observes that the associated covariant derivative $\frac{D}{Dt} [g(t),\xi(t)]_G$ of a curve $[g(t),\xi(t)]_G$ in $\operatorname{Ad}G$ corresponds, via the isomorphism \eqref{Adjoint_bundle_identification}, to the ordinary time derivative in $ \mathfrak{g}  $:
\[
\frac{d}{dt} \eta (t)= \frac{d}{dt} \left( \operatorname{Ad}_{g(t) }\xi(t) \right) ,\quad\text{resp.} \quad \frac{d}{dt} \eta (t)=\frac{d}{dt}\left(  \operatorname{Ad}_{g(t)^{-1} }\xi(t) \right) .
\]
Therefore, in the case $Q=G$, the bundle isomorphism $ \alpha _ \mathcal{A} ^{ (k) }: \left( T^{ (k) }G \right) /G\rightarrow k \operatorname{Ad}G \simeq k \mathfrak{g}  $ becomes
\begin{equation}\label{alpha_Lie_group_right}
\alpha_k\left(\left[\left[g \right]_{g_0}^{(k)} \right]_G \right): = 
\left(\dot{g}(0) g(0) ^{-1}, \left.\frac{d}{dt}\right|_{t=0}\dot{g}(t) g(t)^{-1}, 
\ldots, \left.\frac{d^{k-1}}{dt^{k-1}}\right|_{t=0}\dot{g}(t) g(t)^{-1} \right),
\end{equation}
respectively,
\begin{equation}\label{alpha_Lie_group_left}
\alpha_k\left(\left[\left[g \right]_{g_0}^{(k)} \right]_G \right): = 
\left(g(0) ^{-1}\dot{g}(0) , \left.\frac{d}{dt}\right|_{t=0} g(t)^{-1}\dot{g}(t), 
\ldots, \left.\frac{d^{k-1}}{dt^{k-1}}\right|_{t=0} g(t)^{-1}\dot{g}(t) \right),
\end{equation}
where $k\mathfrak{g}$ denotes the sum of $k$ copies of $\mathfrak{g}$.
Note that in this particular case, one may choose $g_0 =e$ since we have $\left[\left[g \right]_{g_0}^{(k)} \right]_G=\left[\left[g g_0\right]_{e}^{(k)} \right]_G$. The reduced Lagrangian is thus a map $\ell=\ell(\xi,...,\xi^{(k-1)}):k\mathfrak{g}\rightarrow\mathbb{R}$.

\subsection{Ostrogradsky momenta and higher order Hamilton equations}\label{Ostrogradsky_section}

It is well-known that the Hamiltonian formulation of the $k^{th}$-order Euler-Lagrange equations is obtained via the \textit{Ostrogradsky momenta} defined locally by
\[
p_{(i)}:=\sum_{j=0}^{k-i-1}(-1)^j\frac{d^j}{d t^j}\frac{\partial L}{\partial q^{(i+j+1)}},\quad i=0,...,k-1.
\]
So, for example, we have
\[
p_{(0)}=\frac{\partial L}{\partial \dot q}-\frac{d }{d t}\frac{\partial L}{\partial \ddot q}+...+(-1)^{k-1}\frac{d ^{k-1}}{d t^{k-1}}\frac{\partial L}{\partial q^{(k)}}\quad\text{and}\quad p_{(k-1)}=\frac{\partial L}{\partial q^{(k)}}.
\]
These momenta are encoded in the \textit{Ostrogradsky-Legendre transform} $\mathsf{Leg}:T^{(2k-1)}Q\rightarrow T^*\left(T^{(k-1)}Q\right)$ that reads locally $\mathsf{Leg}\left(q,\dot q,..., q^{(2k-1)}\right)=\left(q,...,q^{(k-1)},p_{(0)},...,p_{(k-1)}\right)$. We refer to  \cite{dLRo1985} for the intrinsic definition of the Legendre transform as well as for the geometric formulation of higher order Lagrangian dynamics. In the same way as in the first order case, the \textit{Poincar\'e-Cartan forms} associated to $L$ are defined by
\[
\Theta_L:=\mathsf{Leg}^*\Theta_{can}\in\Omega^1\left(T^{(2k-1)}Q\right)\quad\text{and}\quad \Omega_L:=\mathsf{Leg}^*\Omega_{can}\in\Omega^2\left(T^{(2k-1)}Q\right),
\]
where $\Theta_{can}$ and $\Omega_{can}$ are the canonical forms on $T^*\left(T^{(k-1)}Q)\right)$. The energy function $E_L:T^{(2k-1)}Q\rightarrow\mathbb{R}$ associated to $L$ is defined by
\[
E_L\left([q]^{(2k-1)}\right)=\left\langle\mathsf{Leg}\left([q]^{(2k-1)}\right),[q^{(k-1)}]^{(1)}\right\rangle-L\left([q]^{(k)}\right),
\]
where the bracket denotes the duality pairing between $T\left(T^{(k-1)}Q\right)$ and $T^*\left(T^{(k-1)}Q\right)$. Locally we have
\[
\Theta_L=\sum_{i=0}^{k-1}p_{(i)}dq^{(i)},\quad \Omega_L=\sum_{i=0}^{k-1}dq^{(i)}\wedge dp_{(i)},\quad E_L=\sum_{i=0}^{k-1}p_{(i)}q^{(i+1)}-L(q,...,q^{(k)}).
\]
The Lagrangian $L$ is said to be \textit{regular} if $\Omega_L$ is a symplectic form or, equivalently, if $\mathsf{Leg}$ is a local diffeomorphism. In this case the solution of the $k^{th}$-order Euler-Lagrange are the integral curves of the \textit{Lagrangian vector field} $X_L\in\mathfrak{X}\left(T^{(2k-1)}Q\right)$ defined by
\[
\mathbf{i}_{X_L}\Omega_L=\mathbf{d}E_L.
\]
When $\mathsf{Leg}$ is a global diffeomorphism, then $L$ is \textit{hyperregular} and the associated Hamiltonian is defined by
\[
H:=E_L\circ\mathsf{Leg}^{-1}:T^*\left(T^{(k-1)}Q\right)\rightarrow\mathbb{R}
\]
In this case, the $k^{th}$-order Euler-Lagrange equations are equivalent to the canonical Hamilton equations associated to $H:T^*\left(T^{(k-1)}Q\right) \rightarrow \mathbb{R}$ and are locally given by
\[
\frac{d}{dt}q^{(i)}=\frac{\partial H}{\partial p_{(i)}},\quad \frac{d}{dt}p_{(i)}=-\frac{\partial H}{\partial q^{(i)}},\quad i=0,...,k-1.
\]
The \textit{solution} is the integral curve of the Hamiltonian vector field $X_H\in\mathfrak{X}\left(T^*\left(T^{(k-1)}Q\right)\right)$ defined by 
\[
\mathbf{i}_{X_H}\Omega_{can}=\mathbf{d}H.
\]

\section{Higher order Euler-Poincar\'e reduction}\label{higher_order_EP_sec}

In this section by following \cite{GBHoMeRaVi2011}, we derive the $k^{th}$-order Euler-Poincar\'e equations by reducing the variational principle associated to a right (resp. left) $G$-invariant Lagrangian $L=L(g, \dot g, ..., g^{ (k) }): T^ {(k)} Q\rightarrow \mathbb{R} $ in the special case when the configuration manifold $Q$ is the Lie group $G$ and the action is by right (resp. left) multiplication.

\paragraph{Constrained variations.} As we have seen, the reduced Lagrangian $\ell= \ell ( \xi_1, \xi  _2 ,..., \xi _k ): k \mathfrak{g}  \rightarrow \mathbb{R} $ is induced by the quotient map $T^{(k)}G \rightarrow T^ { (k) } G/G = k \mathfrak{g}$ given by 
\begin{equation}\label{quotmap}
( g, \dot g,..., g^{(k)}) \mapsto  \left( \xi,..., 
\xi^{ (k-1) } \right) ,\quad \text{where}\quad 
\xi := \dot g g ^{-1},\quad\text{resp.}\quad  
\xi := g ^{-1} \dot g 
\end{equation}
and obtained by particularizing the quotient map of \eqref{quotmapLP}.
The variations of the quantities in the quotient map (\ref{quotmap}) are thus given by
\begin{align}\label{constr_var}
\delta \xi  &= \frac{d}{dt}\eta \mp [ \xi, \eta  ]\nonumber\\
\delta \dot \xi &=  \frac{d}{dt} \delta \xi\nonumber\\
\delta \ddot \xi &=  \frac{d^2}{dt^2} \delta \xi  \\
&......\nonumber\\
\delta \frac{d^j}{dt^j}\xi &=  \frac{d^j}{dt^j} \delta \xi,\nonumber
\end{align} 
where $ \eta = (\delta g) g^{-1} $, resp. 
$ \eta =g^{-1} (\delta g)$ for right, resp. left invariance. Therefore, the variations $ \eta \in \mathfrak{g}  $ are such that $\eta ^{ (j) }$ vanish at the endpoints, for all $j=0,..., k-1$.

\paragraph{Hamilton's principle.}
The Euler-Poincar\'e equations for the reduced Lagrangian $\ell$  follow from Hamilton's principle $\delta S = 0$ with $S=\int \ell\,dt$ by using these variations, as
\begin{align*}
\delta \int_{t _1 } ^{ t _2 }\ell \left( \xi, \dot \xi ,..., \xi ^{ (k-1) }\right) dt 
&=\sum _{ j=0}^ {k-1} \int_{t _1 } ^{ t _2 }\left\langle \frac{\delta \ell }{\delta  \xi^{ (j) }}, \delta  \xi^{ (j) } \right\rangle dt
= \sum _{ j=0}^ {k-1} \int_{t _1 } ^{ t _2 }\left\langle \frac{\delta \ell }{\delta  \xi^{ (j) } }, \frac{d^j}{dt^j}\delta  \xi  \right\rangle dt\\
&=\sum _{ j=0}^ {k-1} \int_{t _1 } ^{ t _2 }\left\langle (-1)^j \frac{d^j}{dt^j}\frac{\delta \ell }{\delta  \xi^{ (j) } },  \delta  \xi  \right\rangle dt\\
&= \int_{t _1 } ^{ t _2 }\left\langle \sum _{ j=0}^ {k-1} (-1)^j \frac{d^j}{dt^j}\frac{\delta \ell }{\delta  \xi^{ (j) } }, \frac{d}{dt}\eta \mp [ \xi , \eta  ]  \right\rangle dt\\
&= \int_{t _1 } ^{ t _2 }\left\langle \left( - \frac{d}{dt} \mp \operatorname{ad}^*_ \xi \right)  \sum _{ j=0}^ {k-1} (-1)^j \frac{d^j}{dt^j}\frac{\delta \ell }{\delta  \xi^{ (j) } },  \eta   \right\rangle dt\,,
\end{align*} 
and applying the vanishing endpoint conditions when integrating by parts. 
Therefore, stationarity $(\delta S = 0)$ implies the {\bfi k$^{\text{\bfi th}}$-order Euler-Poincar\'e equations},
\begin{framed}
\begin{equation}\label{EP_k}
\left(\frac{d}{dt}  \pm \operatorname{ad}^*_ \xi \right)  
\sum _{ j=0}^ {k-1} (-1)^j \frac{d^j}{dt^j}  \frac{\delta \ell }{\delta  \xi^{ (j) } }=0.
\end{equation}
\end{framed}

\noindent Formula (\ref{EP_k}) takes the following forms for various choices of $k=1,2,3$:\\

\noindent
If $k=1$:
\[
\left( \frac{d}{dt} \pm \operatorname{ad}^*_ \xi \right) \frac{\delta \ell }{\delta  \xi}=0
,\]
If $k=2$:
\begin{equation}\label{EP_2}
\left( \frac{d}{dt}\pm \operatorname{ad}^*_ \xi \right) 
\left( \frac{\delta  \ell}{\delta \xi } - \frac{d}{dt} \frac{\delta \ell }{\delta \dot \xi } \right) =0
,\end{equation}
If $k=3$:
\[
\left(\frac{d}{dt}  \pm \operatorname{ad}^*_ \xi \right) \left( \frac{\delta  \ell}{\delta \xi } 
- \frac{d}{dt} \frac{\delta \ell }{\delta \dot \xi } 
+ \frac{d^2}{dt^2}\frac{\delta \ell }{\delta \ddot \xi }\right) =0
.\]
The first of these is the usual Euler-Poincar\'e equation. The others adopt a \emph{factorized} form in which the Euler-Poincar\'e operator $(d/dt \pm \operatorname{ad}^*_ \xi )$ is applied to the Euler-Lagrange operation on the reduced Lagrangian $\ell ( \xi, \dot \xi, \ddot{\xi},... ) $ at the given order.

The results obtained above are summarized in the following theorem.

\begin{theorem}[$k^{th}$-order Euler-Poincar\'e reduction] Let $L:T^{(k)}G \rightarrow \mathbb{R}$ be a $G$-in\-va\-riant Lagrangian and $\ell: k \mathfrak{g}  \rightarrow \mathbb{R} $ the associated reduced Lagrangian. Let $g(t)$ be a curve in $G$ and $ \xi (t)= \dot g(t)g(t)^{-1}$, resp. $ \xi (t)= g(t)^{-1}\dot g(t)$ be the reduced curve in the Lie algebra $ \mathfrak{g} $. Then the following assertions are equivalent.
\begin{itemize}
\item[\rm (i)] The curve $g(t)$ is a solution of the 
$k^{th}$-order Euler-Lagrange equations for $L$.
\item[\rm (ii)] Hamilton's variational principle
\[\delta \int_{t _0 }^{t _1 } L\left( g, \dot g, ..., g^{ (k) } \right) dt =0
\]
holds using variations $ \delta g$ such that $\delta g ^{ (j) }$ vanish at the endpoints for $j=0,...,k-1$.
\item[\rm (iii)] The $k^{th}$-order Euler-Poincar\'e equations for $\ell:k \mathfrak{g}\rightarrow\mathbb{R}$
hold:
\begin{equation}
\left(\frac{d}{dt} \pm \operatorname{ad}^*_ \xi \right)  
\sum _{ j=0}^ {k-1} (-1)^j \frac{d^j}{dt^j}  \frac{\delta \ell }{\delta  \xi^{ (j) } }=0.
\end{equation}
\item[\rm (iv)] The constrained variational principle
\[
\delta \int_{t _0 }^{t _1 } \ell \left(  \xi , \dot \xi ,..., \xi ^ { (k) } \right) =0
\]
holds for constrained variations $ \delta \xi = \frac{d}{dt} \eta \mp [ \xi, \eta  ]$, where $ \eta $ is an arbitrary curve in $\mathfrak{g}  $ such that $\eta ^{ (j) }$ vanish at the endpoints, for all $j=0,..., k-1$.
\end{itemize}
\end{theorem}

\medskip

We now quickly recall from \cite{GBHoMeRaVi2011} how the higher order Euler-Poincar\'e theory applies to geodesic splines on Lie groups.

\paragraph{Example: Riemannian 2-splines on Lie groups.}
Fix a right, resp. left invariant Riemannian metric $\gamma $ on the Lie group $G$ and let $\|\cdot\|^2$ be its corresponding squared metric.  Consider the Lagrangian $L: T^ { (2)} G\rightarrow \mathbb{R}$ for Riemannian $2$-splines, given by
\begin{equation}
\label{un-reduced-lag-ell}
L(g, \dot{ g}, \ddot{ g})=\frac{1}{2} \left \| \frac{D}{Dt}\dot g\right\|^2,
\end{equation}
in which $D/Dt$ denotes covariant derivative in time. 
It is shown in \cite{GBHoMeRaVi2011} that the reduced Lagrangian $\ell: 2 \mathfrak{g}  \rightarrow \mathbb{R}$ associated to $L$ is given by 
\begin{equation}
\label{reduced-lag-ell}
\ell( \xi , \dot{ \xi })= \frac{1}{2}\left \| \dot{ \xi }^\flat \pm \operatorname{ad}^*_ \xi \xi^\flat \right \| ^2,
\end{equation}
where $\flat:\xi\in \mathfrak{g}\mapsto \gamma_e(\xi, \cdot ) \in \mathfrak{g}^*$ is the flat operator associated to $\gamma$.
From formula \eqref{EP_2} with $k=2$ one then finds the $2^{nd}$-order Euler-Poincar\'e equation
\begin{equation}
\label{2nd-EPeqns1}
\left(\frac{d}{dt} \pm \operatorname{ad}^*_ \xi \right) \left(\frac{d}{dt}  \eta^\flat 
\pm  \operatorname{ad}^*_{ \eta} \xi ^\flat \pm \left(\operatorname{ad}_{\eta} \xi \right)^\flat\right) =0
, \quad \hbox{with}\quad
 \eta^\flat := \dot{ \xi }^\flat \pm \operatorname{ad}^* _ \xi \xi^\flat.
\end{equation}
If the metric is both left and right invariant (bi-invariant) further simplifications arise. Indeed, in this case we have $\operatorname{ad}^* _ \xi \xi^\flat=0$ so that $\eta= \dot{ \xi}$ and  the equations in \eqref{2nd-EPeqns1} become
\begin{equation}
\left(\frac{d}{dt}\pm \operatorname{ad}^*_ \xi \right) \ddot{\xi}\,^\flat =0\quad\text{or}\quad \dddot{\xi}\mp \left[\xi , \ddot{ \xi }\right]=0,
\label{CrSLe-commutator}
\end{equation}
as in \cite{CrSL1995}. Note that in this case, the reduced Lagrangian \eqref{reduced-lag-ell} is given simply by $\ell( \xi , \dot \xi )=\frac{1}{2} \|\dot \xi \|^2$.

\section{Higher order Lagrange-Poincar\'e reduction}\label{Lagrange_Poincare_section}

Here we generalize the method of Lagrange-Poincar\'e reduction in \cite{CeMaRa2001} to higher order $G$-invariant Lagrangians defined on $T^{(k)}Q$. Recall from \S\ref{kth_order_tgt_bundles} that we consider a free and proper right (resp. left) action $\Phi$ of $G$ on $Q$ and its lift $\Phi^{(k)}$ on the $k^{th}$-order tangent bundle $T^{(k)}Q$. By fixing a principal connection $\mathcal{A}$ on the principal bundle $\pi:Q\rightarrow Q/G$, the quotient space $\left(T^{(k)}Q\right)/G$ can be identified with the bundle $T^{ (k) }(Q/G)\oplus k \operatorname{Ad}Q$.

\paragraph{Constrained variations.} The main departure point is to compute the constrained variations of
\begin{equation}\label{LP_reduced_variables}
\left( \rho , \dot \rho , ..., \rho ^{ (k)}, \sigma , \dot \sigma , ..., \sigma ^{ (k-1)} \right)= \alpha _ \mathcal{A} ^ {(k) } \left( \left[[ q]^{ (k) }_{ q _0 }\right] _G \right)\in T^{ (k) }(Q/G)\oplus k \operatorname{Ad}Q
\end{equation}
induced by a variation $ \delta q(t)= \left.\frac{d}{ds}\right|_{s=0} q(t,s)$ of the curve $q(t)\in Q$. Since
\[
( \rho , \dot \rho ,..., \rho ^{ (k) })= T^{ (k) }\pi \left([ q]^{ (k) }_{q _0 } \right) = [ \pi\circ q]_{\rho}^{ (k) },
\]
the variations $ \delta \rho $ of $ \rho $ are arbitrary except for the endpoint conditions $ \delta \rho^{ (j) } (t_i)=0$, for all $i=1,2$, $j=1,.., k-1$. The variations of $ \sigma ^j $ may be computed with the help of a fixed connection $ \mathcal{A} $. For $\sigma(t):=[q(t), \mathcal{A}( \partial _tq(t))]_G\in \operatorname{Ad}Q$, we have
\begin{align*} 
\delta \sigma (t)&= \left.\frac{D}{Ds}\right|_{s=0} \sigma(t,s)= \left.\frac{D}{Ds}\right|_{s=0} \left[ q(t,s), \mathcal{A} ( \partial _t q(t,s)) \right]_G\\
&\stackrel{\eqref{cov_der_general}}
=\left [ q(t), \left.\frac{d}{ds}\right|_{s=0}  \mathcal{A} ( \partial _t q(t,s)) \pm \left[\mathcal{A} \left( \delta  q(t)\right), \mathcal{A} ( \partial _t q(t))\right]\right]_G\\
&=\left [ q(t), \frac{d}{dt} \mathcal{A} ( \delta q(t))  +\mathbf{d} \mathcal{A} \left(\delta q(t), \partial _t q(t)\right) \pm \left[\mathcal{A} \left( \delta  q(t)\right), \mathcal{A} ( \partial _t q(t))\right]\right]_G\\
&=\left [ q(t), \frac{d}{dt} \mathcal{A} ( \delta q(t))  +\mathcal{B} \left(\delta q(t), \partial _t q(t)\right) \right]_G\\
&= \frac{D}{Dt}[q(t), \mathcal{A} ( \delta q(t))]_G \mp \left[ q(t), [ \mathcal{A} ( \partial _t  q), \mathcal{A} ( \delta q(t))]\right]_G+ \left[ q(t), \mathcal{B} \left(\delta q(t), \partial _t q(t)\right) \right]_G.
\end{align*}
This computation implies
\begin{equation}
\delta \sigma (t)
=
 \frac{D}{Dt}\eta (t) \mp [ \sigma (t), \eta (t)] + \tilde{\mathcal{B}} (\delta \rho (t), \dot \rho (t))
\label{delta-sigma}
\end{equation}
for $\eta(t):=[q(t), \mathcal{A} ( \delta q(t))]_G \in
\operatorname{Ad}Q $ and where $\mathcal{B}:=\mathbf{d}\mathcal{A}\pm[\mathcal{A},\mathcal{A}]\in\Omega^2(Q,\mathfrak{g})$ is the \textit{curvature $2$-form} and $\tilde{\mathcal{B}}\in\Omega^2(Q/G,\operatorname{Ad}Q)$ is the \textit{reduced curvature}. Our conventions are $[\mathcal{A}, \mathcal{A}](u,v):= [\mathcal{A}(u), \mathcal{A}(v)]$ for all
$u,v \in T_qQ$.

For $ \dot \sigma $, using the formula
\[
\frac{D}{Dt}\frac{D}{Ds} \sigma (s,t)-  \frac{D}{Ds}\frac{D}{Dt} \sigma (s,t)=\pm\left[\tilde{\mathcal{B}}( \dot \rho, \delta \rho ), \sigma \right]
\]
leads to
\begin{align*}
\delta\dot \sigma  (t)&= \left.\frac{D}{Ds}\right|_{s=0} \dot \sigma (s,t)=\left.\frac{D}{Ds}\right|_{s=0}\frac{D}{Dt}\sigma (s,t)=\frac{D}{Dt}\left.\frac{D}{Ds}\right|_{s=0} \sigma (s,t)\mp \left[\tilde{\mathcal{B}}( \dot \rho, \delta \rho ), \sigma  \right]\\
&= \frac{D}{Dt}\delta  \sigma  (s,t)\mp \left[\tilde{\mathcal{B}}( \dot \rho, \delta \rho ), \sigma  \right].
\end{align*} 
Then, similarly
\begin{align*}
\delta \ddot \sigma  (t)&= \left.\frac{D}{Ds}\right|_{s=0} \ddot \sigma  (s,t)=\left.\frac{D}{Ds}\right|_{s=0}\frac{D}{Dt}\dot \sigma  (s,t)= \frac{D}{Dt}\delta  \dot{ \sigma} (s,t)\mp \left[\tilde{\mathcal{B}}( \dot \rho, \delta \rho ), \dot{ \sigma }  \right]\\
&= \frac{D^2}{Dt^2}\delta  \sigma(s,t)\mp \frac{D}{Dt} \left[\tilde{\mathcal{B}}( \dot \rho, \delta \rho ), \sigma  \right] \mp \left[\tilde{\mathcal{B}}( \dot \rho, \delta \rho ), \dot{ \sigma}  \right].
\end{align*} 
In general, one finds
\begin{align}
\delta \sigma ^{(j)} (t)&= \left.\frac{D}{Ds}\right|_{s=0} \sigma ^{ (j) } (s,t)= \frac{D^j}{Dt^j}\delta  \sigma (s,t)\mp \sum _{p=0}^{j-1} \frac{D^p}{Dt^p} \left[\tilde{\mathcal{B}}( \dot \rho, \delta \rho ), \sigma ^{ (j-1-p) } \right].
\label{djdot-sig}
\end{align}

\paragraph{Hamilton's principle.}
We can now compute the $k^{th}$-order Lagrange-Poincar\'e equations. Using Hamilton's principle we have
\begin{align*}
&\delta \int_{ t_1 }^{ t_2 } \ell \left( \rho , \dot\rho ,..., \rho ^ { (k) }, \sigma , \dot \sigma ,..., \sigma ^{ (k-1) } \right) dt \\
&\qquad = \sum_{j=0}^{k} \int_ { t _1 } ^{ t _2 } \left\langle \frac{\delta \ell }{\delta  \rho ^ { (j) }} , \delta \rho ^{ (j) } \right\rangle dt  +\sum_{j=0}^{k-1} \int_ { t _1 } ^{ t _2 } \left\langle \frac{\delta \ell }{\delta  \sigma ^ { (j) }} , \delta \sigma  ^{ (j) } \right \rangle dt.
\end{align*} 
The first term produces the usual expression 
\[
\int_ { t _1 } ^{ t _2 } \left\langle \sum_{j=0}^{k} (-1)^j\frac{d^j}{dt^j}  \frac{\delta \ell }{\delta  \rho ^ { (j) }}, \delta \rho  \right\rangle dt.
\]
For the second term, we have
\begin{align*} 
 &\int_ { t _1 } ^{ t _2 } \left\langle \frac{\delta \ell }{\delta  \sigma ^ { (j) }} , \delta \sigma  ^{ (j) } \right \rangle dt \\
 &= \int_ { t _1 } ^{ t _2 } \left\langle \frac{\delta \ell }{\delta  \sigma ^ { (j) }} ,\frac{D^j}{Dt^j}\delta  \sigma \right\rangle dt \mp  \sum _{p=0}^{j-1} \int_ { t _1 } ^{ t _2 } \left\langle \frac{\delta \ell }{\delta  \sigma ^ { (j) }} ,\frac{D^p}{Dt^p} \left[\tilde{\mathcal{B}}( \dot \rho, \delta \rho ), \sigma ^{ (j-1-p) } \right] \right\rangle dt \\
 &= \int_ { t _1 } ^{ t _2 } \left\langle (-1)^j \frac{D^j}{Dt^j}\frac{\delta \ell }{\delta  \sigma ^ { (j) }} ,\delta  \sigma \right\rangle dt \mp  \sum _{p=0}^{j-1} \int_ { t _1 } ^{ t _2 } \left\langle(-1)^p  \frac{D^p}{Dt^p} \frac{\delta \ell }{\delta  \sigma ^ { (j) }} ,\left[\tilde{\mathcal{B}}( \dot \rho, \delta \rho ), \sigma ^{ (j-1-p) } \right] \right\rangle dt \\
	 &= \int_ { t _1 } ^{ t _2 } \left\langle (-1)^j \frac{D^j}{Dt^j}\frac{\delta \ell }{\delta  \sigma ^ { (j) }} ,\delta  \sigma \right\rangle dt \pm  \sum _{p=0}^{j-1} \int_ { t _1 } ^{ t _2 } \left\langle (-1)^p \operatorname{ad}^*_{ \sigma ^{(j-1-p)} }\frac{D^p}{Dt^p} \frac{\delta \ell }{\delta  \sigma ^ { (j) }} ,\tilde{\mathcal{B}}( \dot \rho, \delta \rho )\right\rangle dt 
\end{align*}
upon dropping endpoint terms. We then insert expression (\ref{delta-sigma}) for $ \delta \sigma $ in the
first summand which becomes
\[
\int_ { t _1 } ^{ t _2 }\left( \left\langle \left( - \frac{D}{Dt} \mp \operatorname{ad}^*_ \sigma \right)  (-1)^j \frac{D^j}{Dt^j}\frac{\delta \ell }{\delta  \sigma ^ { (j) }} ,\eta  \right\rangle - \left\langle \left\langle (-1)^j \frac{D^j}{Dt^j}\frac{\delta \ell }{\delta  \sigma ^ { (j) }} , \mathbf{i} _{ \dot{ \rho }} \tilde{\mathcal{B}}  \right\rangle, \delta \rho \right\rangle \right)dt.
\]
We thus obtain the system of {\bfi k$^{\text{\bfi th}}$-order Lagrange-Poincar\'e equations}
\begin{framed} 
\begin{equation}\label{Lagrange_poincare_eq}
\!\!\!\!\!\!\!\!\!\left\{
\begin{array}{l}
\displaystyle \vspace{0.2cm}
\sum_{j=0}^{k} (-1)^j\frac{d^j}{dt^j}  \frac{\delta \ell }{\delta  \rho ^ { (j) }}
=
 \left\langle \sum_{j=0}^{k-1}\left((-1)^j  
 \frac{D^j}{Dt^j}\frac{\delta \ell }{\delta \sigma^{(j)}}  \mp \sum_{p=0}^{j-1} (-1)^p \operatorname{ad}^*_{ \sigma ^{(j-1-p)} }\frac{D^p}{Dt^p} \frac{\delta \ell }{\delta  \sigma ^ { (j) }}\right)
 ,\mathbf{i} _{\dot \rho}\tilde{\mathcal{B}}\right\rangle
,\\
\displaystyle\left(\frac{D}{Dt} \pm \operatorname{ad}^*_ \sigma \right)  \sum_{j=0}^{k-1}(-1)^j \frac{D^j}{Dt^j}\frac{\delta \ell }{\delta  \sigma ^ { (j) }}=0.
\end{array}
\right.
\end{equation} 
\end{framed}

These equations recover the higher order Euler-Poincar\'e equations \eqref{EP_k} in the special case $Q=G$.
For $k=1$, \eqref{Lagrange_poincare_eq} reduces to
\begin{equation} 
\left\{
\begin{array}{l}
\displaystyle \vspace{0.2cm} \frac{\delta \ell }{\delta  \rho} - \frac{d}{dt} \frac{\delta \ell }{\delta  \dot{\rho}}=\left\langle \frac{\delta \ell }{\delta  \sigma },\mathbf{i} _{\dot \rho}\tilde{\mathcal{B}}\right\rangle
\\
\displaystyle\left(\frac{D}{Dt} \pm \operatorname{ad}^*_ \sigma \right) \frac{\delta \ell }{\delta  \sigma }=0
\end{array}
\right.
\end{equation}
in agreement with the Lagrange-Poincar\'e equations obtained in \cite{CeMaRa2001}. For 
$k=2$ we get
\begin{equation}\label{Lagr_Poinc_2}
\left\{
\begin{array}{l}
\displaystyle \vspace{0.2cm} 
\frac{\delta \ell }{\delta  \rho} - \frac{d}{dt} \frac{\delta \ell }{\delta  \dot{\rho}}
+ \frac{d^2}{dt^2} \frac{\delta \ell }{\delta  \ddot{\rho}} =\left\langle \frac{\delta \ell }{\delta  \sigma },\mathbf{i} _{\dot \rho}\tilde{\mathcal{B}}\right\rangle
+ \left\langle - \frac{D}{Dt} \frac{\delta \ell }{\delta  \dot{\sigma }}\mp \operatorname{ad}^*_\sigma \frac{\delta \ell }{\delta  \dot{ \sigma }},\mathbf{i} _{\dot \rho}\tilde{\mathcal{B}}\right\rangle
,\\
\displaystyle\left(\frac{D}{Dt} \pm \operatorname{ad}^*_ \sigma \right)
\left( \frac{\delta \ell }{\delta  \sigma } 
-  \frac{D}{Dt}\frac{\delta \ell }{\delta  \dot{\sigma} }  \right)
 =0.
\end{array}
\right.
\end{equation}

\paragraph{Case of zero curvature.}
Note that when the curvature vanishes, then the Lagrange-Poincar\'e equations are simply given by the $k^{th}$-order Euler-Lagrange equations for the variables $( \rho , \dot \rho , ..., \rho ^{ (k) })$ together with the $k^{th}$-order Euler-Poincar\'e equations in the variables $\left(  \sigma , \dot \sigma ,..., \sigma ^{(k-1) } \right) $.

\medskip

The results obtained above are summarized in the following theorem.

\begin{theorem}[$k^{th}$-order Lagrange-Poincar\'e reduction] Consider a free and proper action of a Lie group $G$ on a manifold $Q$ and let $L:T^{(k)}Q \rightarrow \mathbb{R}  $ be a $G$-invariant Lagrangian. Choose a principal connection $ \mathcal{A} \in \Omega^1(Q; \mathfrak{g})$ on the principal bundle $\pi: Q\rightarrow Q/G$ and let $\ell: T^{ (k) }(Q/G)\oplus k \operatorname{Ad}Q  \rightarrow \mathbb{R} $ be the associated reduced Lagrangian. Let $q(t)$ be a curve in $Q$ and let $\left( \rho , \dot \rho , ..., \rho ^{ (k)}, \sigma , \dot \sigma , ..., \sigma ^{ (k-1)} \right)$ be the reduced curves obtained via the relation \eqref{LP_reduced_variables}. Then the following assertions are equivalent.
\begin{itemize}
\item[\rm (i)] The curve $q(t)$ is a solution of the $k^{th}$-order Euler-Lagrange equations for $L$.
\item[\rm (ii)] Hamilton's variational principle
\[\delta \int_{t _0 }^{t _1 } L\left( q, \dot q, ..., q^{ (k) } \right) dt =0
\]
holds using variations $ \delta q$ such that $\delta q^{ (j) }$ vanish at the endpoints for $j=0,...,k-1$.
\item[\rm (iii)] The $k^{th}$-order Lagrange-Poincar\'e equations for $\ell: T^{ (k) }(Q/G)\oplus 
k \operatorname{Ad}Q \rightarrow \mathbb{R}$ hold:
\[
\left\{
\begin{array}{l}
\displaystyle \vspace{0.2cm}\sum_{j=0}^{k} (-1)^j\frac{d^j}{dt^j}  \frac{\delta \ell }{\delta  \rho ^ { (j) }}= \sum_{j=0}^{k-1} \left\langle (-1)^j \frac{D^j}{Dt^j}\frac{\delta \ell }{\delta  \sigma ^ { (j) }}  \mp \sum_{p=0}^{j-1} (-1)^p \operatorname{ad}^*_{ \sigma ^{(j-1-p)} }\frac{D^p}{Dt^p} \frac{\delta \ell }{\delta  \sigma ^ { (j) }} ,\mathbf{i} _{\dot \rho}\tilde{\mathcal{B}}\right\rangle
,\\
\displaystyle\left(\frac{D}{Dt} \pm \operatorname{ad}^*_ \sigma \right)  \sum_{j=0}^{k-1}(-1)^j \frac{D^j}{Dt^j}\frac{\delta \ell }{\delta  \sigma ^ { (j) }}=0.
\end{array}
\right.
\]
\item[\rm (iv)] The constrained variational principle
\[
\delta \int_{t _0 }^{t _1 } \ell \left( \rho , \dot \rho , ..., \rho ^{ (k)}, \sigma , \dot \sigma , ..., \sigma ^{ (k-1)} \right) =0
\]
holds for variations $ \delta \rho $ of $ \rho $ with the endpoint conditions $ \delta \rho^{ (j) } (t_i)=0$, for all $i=1,2$, $j=0,.., k-1$ and for constrained variations
\[
\delta \sigma ^{(j)}=\frac{D^j}{Dt^j}\delta  \sigma \mp \sum _{p=0}^{j-1} \frac{D^p}{Dt^p} \left[\tilde{\mathcal{B}}( \dot \rho, \delta \rho ), \sigma ^{ (j-1-p) } \right]\quad\text{with}\quad 
\delta\sigma=\frac{D}{Dt}\eta\mp [ \sigma , \eta ] + \tilde{\mathcal{B}} (\delta \rho , \dot \rho ),
\]
where $ \eta $ is an arbitrary curve in $\operatorname{Ad}Q$ such that $\frac{D^j}{Dt^j}\eta$ vanish at the endpoints, for all $j=0,..., k-1$.
\end{itemize}
\end{theorem}

\paragraph{Example: Wong's equations.} A standard example of $G$-invariant Lagrangian in the context of a principal bundle $Q\rightarrow M=Q/G$ is the \textit{Kaluza-Klein} Lagrangian
\[
L(u_q)=\frac{1}{2}\gamma_{\pi(q)}(T\pi(u_q),T\pi(u_q))+\frac{1}{2}\kappa\left(\mathcal{A}(u_q),\mathcal{A}(u_q)\right),
\]
where $\gamma$ is a Riemannian metric on $M$, $\kappa$ is an Ad-invariant inner product on the Lie algebra $\mathfrak{g}$, and $\mathcal{A}$ is a principal connection. The reduced Lagrangian induced on $TQ/G\simeq TM\oplus\operatorname{Ad}Q$ reads
\[
\ell(\rho,\dot\rho,\sigma)=\frac{1}{2}\gamma_\rho(\dot\rho,\dot\rho)+\frac{1}{2}\bar\kappa(\sigma,\sigma),
\]
where $\bar\kappa$ is the bundle metric induced by $\kappa$ on $\operatorname{Ad}Q$. The Lagrange-Poincar\'e equations associated to this Lagrangian are thus given by
\[
\frac{D}{Dt}\,\dot\rho=-\left\langle\bar\mu,\mathbf{i}_{\dot\rho}\tilde{\mathcal{B}}\right\rangle,\quad \frac{D}{Dt}\bar\mu=0,
\]
where $\bar\mu:=\bar\kappa(\sigma,\_\,)\in
\operatorname{Ad}^*Q$, \cite{CeMaRa2001}. These equations 
are known as \textit{Wong's equations} and arise  in at 
least two different interesting contexts. The first of 
these, in the work of \cite{Wo1970}, \cite{St1970}, 
\cite{We1978}, and \cite{Mo1984}, concerns the dynamics 
of a colored particle in a Yang-Mills field. In this 
case, $\bar\mu$ is the generalized charge, the right 
hand side of the first equation represents the 
generalized Lorentz force associated to the magnetic 
field $\tilde{\mathcal{B}}$, and the second equation the 
expresses charge conservation. The second context is 
that of the falling cat theorem of \cite{Mo1990}, 
\cite{Mo1993}.

A natural second order generalization of the Kaluza-Klein Lagrangian would be
\begin{align}
L(q,\dot q,\ddot q)&=\frac{1}{2}  \gamma_{\pi(q)}
(T\pi(\dot{q}),T\pi(\dot{q}))+\frac{\lambda_1^2}{2} \gamma_{\pi(q)} \left(
\frac{D}{Dt}T\pi(\dot{q}), \frac{D}{Dt}T\pi(\dot{q})\right)
\nonumber\\
&\qquad\qquad\qquad  +  \frac{1}{2}\kappa \left( \mathcal{A}(\dot q),\mathcal{A}(\dot q)\right)+\frac{\lambda_2^2}{2}\kappa \left( \frac{d}{dt}\mathcal{A}(\dot q),\frac{d}{dt}\mathcal{A}(\dot q)\right).
\label{HOKK-Lag}
\end{align}
One easily checks that this Lagrangian is $G$-invariant. Recall from \eqref{cov_der_general} that the covariant derivative of a curve $[q(t),\xi(t)]_G\in\operatorname{Ad}Q$ relative to the principal connection $\mathcal{A}$ is $\frac{D}{Dt}[q(t),\xi(t)]_G=\left[q(t),\dot{\xi} (t)\pm [\mathcal{A} ( \dot q(t)), \xi (t) ]\right] _G$. So, in the particular case $\xi(t)=\mathcal{A}(\dot q(t))$, we have
\[
\frac{D}{Dt}\left[q(t),\mathcal{A}(\dot q(t))\right]_G=\left[q(t), \frac{d}{dt}\mathcal{A}(\dot q(t))\right]_G.
\]
This shows that the associated reduced Lagrangian is 
\[
\ell(\rho,\dot\rho,\ddot\rho,\sigma,\dot\sigma)
=
\frac{1}{2}\gamma_\rho(\dot\rho,\dot\rho)
+\frac{\lambda_1^2}{2}\gamma_\rho\left(\frac{D}{Dt}\dot\rho,\frac{D}{Dt}\dot\rho\right)
+
\frac{1}{2}\bar\kappa(\sigma,\sigma)+\frac{\lambda_2^2}{2}\bar\kappa(\dot\sigma,\dot\sigma).
\]
As in the case of Wong's equations, we define $\bar\mu:=\bar{\kappa}(\sigma,\_\,)\in\operatorname{Ad}^*Q$. We note the formula
\[
\frac{D}{Dt}\bar\mu=\bar{\kappa}\left(\frac{D}{Dt}\sigma,\_\,\right),
\]
where the \emph{same} symbol $D/Dt$ is used to denote the covariant derivative in both $\operatorname{Ad}^*Q$ and $\operatorname{Ad}Q$. We can thus write the functional derivatives as
\[
\frac{\delta\ell}{\delta\sigma}=\bar\mu\quad\text{and}\quad \frac{\delta\ell}{\delta\dot\sigma}=\lambda_2^2\frac{D}{Dt}\bar\mu
\]
From this expression we compute the second order Lagrange-Poincar\'e equations \eqref{Lagr_Poinc_2} as
\begin{equation}\label{2nd_order_Wong}
\left\{
\begin{array}{l}
\vspace{0.2cm}\displaystyle\frac{D}{Dt}\dot \rho (t) -\lambda_1^2\frac{D^3}{Dt^3}\dot \rho(t)- \lambda_1^2R \left( \frac{D}{Dt}\dot \rho(t),\dot \rho(t)
\right) \dot \rho(t)
=
-\left\langle  \left( \bar\mu-\lambda_2^2\frac{D^2}{Dt^2} \bar\mu\mp \operatorname{ad}^*_\sigma \frac{D}{Dt}\bar\mu\right),\mathbf{i} _{\dot \rho}\tilde{\mathcal{B}}\right\rangle
\\
\displaystyle\left(\frac{D}{Dt} \pm \operatorname{ad}^*_ \sigma \right)
\left( \bar\mu
- \lambda_2^2\frac{D^2}{Dt^2}\bar\mu \right)
 =0.
\end{array}
\right.
\end{equation}

\begin{remark}\rm
Recall that the motivation we discussed at the outset for developing the higher-order framework for geometric mechanics involved formulating the boundary value problems needed for the applications of higher-order optimal control methods in, for example, the application of geometric splines in longitudinal data assimilation.  For further discussion of the motivation for developing the higher-order framework for geometric mechanics, additional references and an illustrative finite-dimensional example for template matching on the sphere, see \cite{GBHoMeRaVi2011}.

The higher-order Wong's equations (\ref{2nd_order_Wong}) are relevant from the viewpoint of  
geometric splines, when they are regarded as adding a lower-order ``elastic  
tension" in the higher-order spline equations, rather than being regarded as 
a higher-order deformation of the initial value problem for Wong's equations. 
Indeed, from the viewpoint of the initial value problem, adding the geometric spline terms 
would be regarded as a \emph{singular perturbation} of the lower-order case.
That is, adding lower-order terms is a deformation; adding higher-order  
terms is a singular perturbation. 
Therefore, adding lower-order elastic terms to  
splines is simpler to interpret than adding higher-order terms to  
systems such as Wong's equations that are understood at the lower  
order. This was also the original philosophy in the  
adding ``tension'' to splines in \cite{Schw1966}. Thus, the present section might be fruitfully regarded  
as a deformation of the spline equations, rather than as a singular  
perturbation of Wong's equations. To make this interpretation more explicit, one might 
transfer the $\lambda$-parameters to the Kaluza-Klein terms, rather  
than the higher-order spline terms in the Lagrangian (\ref{HOKK-Lag}).
\end{remark}

\section{Ostrogradsky-Lie-Poisson reduction}\label{Ham-forms-sec}

In this Section we present the Hamiltonian side of the higher-order Euler-Poincar\'e reduction carried out in Section \ref{higher_order_EP_sec}. It consists of a generalization of the Lie-Poisson formulation to higher order, that uses Ostrogradsky momenta defined in \S\ref{Ostrogradsky_section} and encoded in the Legendre transform $\mathsf{Leg}:T^{(2k-1)}G\rightarrow T^*\left(T^{(k-1)}G\right)$. 

\paragraph{Legendre transform.} Given a reduced Lagrangian $\ell:k\mathfrak{g}\rightarrow\mathbb{R}$, the reduced Legendre transform is $\mathsf{leg}:(2k-1)\mathfrak{g}\rightarrow T^*(k-1)\mathfrak{g}\times\mathfrak{g}^*$,
\[
\mathsf{leg}(\xi,\dot\xi,...,\xi^{(2k-1)})=\left(\xi,\dot\xi,...,\xi^{(k-2)},\pi_{(1)},...,\pi_{(k-1)},\pi_{(0)}\right),\quad \pi_{(i)}=\sum_{i=0}^{k-1}(-1)^i\frac{d}{dt}\frac{\delta\ell}{\delta\xi^{(i)}},
\]
so that, for example,
\[
\pi_{(0)}=\frac{\delta\ell}{\delta \xi}-\frac{d}{dt}\frac{\delta\ell}{\delta \dot\xi}+...+(-1)^{k-1}\frac{d^{k-1}}{dt^{k-1}}\frac{\delta\ell}{\delta\xi^{(k-1)}}\quad\text{and}\quad \pi_{(k-1)}=\frac{\delta\ell}{\delta\xi^{(k-1)}}.
\]
As before, notations such as $\frac{d^j}{dt^j}\frac{\delta\ell}{\delta \xi^{(i)}}$ means the expression obtained by developing the formula 
in terms of the independent variables $(\xi,\dot\xi,...,\xi^{(2k-1)})$. The reduced energy associated to $\ell$ is $e_\ell:(2k-1)\mathfrak{g}\rightarrow\mathbb{R}$ given by
\begin{align*}
e_\ell\left(\xi,...,\xi^{(2k-2)}\right)&=\left\langle\mathsf{leg}\left(\xi,...,\xi^{(2k-2)}\right),\left(\xi,...,\xi^{(k-2)},\dot\xi,...,\xi^{(k-1)},\xi\right)\right\rangle-\ell\left(\xi,...,\xi^{(k)}\right)\\
&=\sum_{i=0}^{k-1}\left\langle\pi_{(i)},\xi^{(i)}\right\rangle-\ell\left(\xi,...,\xi^{(k-1)}\right),
\end{align*}
where the bracket in the first line denotes the duality pairing between $T^*(k-1)\mathfrak{g}\times\mathfrak{g}^*$ and $T(k-1)\mathfrak{g}\times\mathfrak{g}$.

\paragraph{Poisson structure.} The reduced Poisson bracket induced on $T^*(k-1)\mathfrak{g}\times\mathfrak{g}^*$ by the canonical symplectic structure on $T^*\left(T^{(k-1)}G\right)$ is the sum of the canonical bracket on $T^*(k-1)\mathfrak{g}$ and the Lie-Poisson structure on $\mathfrak{g}^*$, that is,
\begin{align*}
&\{f,g\}\left(\xi,\dot\xi,...,\xi^{(k-2)},\pi_{(1)},...,\pi_{(k-1)},\pi_{(0)}\right)\\
&\qquad=\{f,g\}_{can}\left(\xi,\dot\xi,...,\xi^{(k-2)},\pi_{(1)},...,\pi_{(k-1)}\right)+\{f,g\}_{\pm}(\pi_{(0)})\\
&\qquad=\sum_{j=1}^{k-1}\left(\frac{\delta f}{\delta \xi^{(j-1)}}\frac{\delta g}{\delta \pi_{(j)}}-\frac{\delta g}{\delta \xi^{(j-1)}}\frac{\delta f}{\delta \pi_{(j)}}\right)\pm\left\langle\pi_{(0)},\left[\frac{\delta f}{\delta \pi_{(0)}},\frac{\delta g}{\delta \pi_{(0)}}\right]\right\rangle.
\end{align*}
The associated {\bfi Ostrogradsky-Lie-Poisson} equations are hence
\begin{framed}
\begin{equation}\label{Ostro_LP_kth}
\left\{\begin{array}{l}
\displaystyle \vspace{0.2cm}\left(\partial _t \pm \operatorname{ad}^*_{\frac{\delta  h}{\delta \pi _{(0)}} }\right)\pi _{(0)} =0
\,,\\
\displaystyle\partial _t \xi^{(j-1)} = \frac{\delta  h}{\delta \pi _{(j)}  },\qquad 
\partial _t \pi _{(j)}  = -\frac{\delta  h}{\delta \xi^{(j-1)} },\qquad j=1,...,k-1,
\end{array}\right.
\end{equation}
\end{framed}

In the hyperregular case, when $h$ is associated to a reduced Lagrangian $\ell$ via the Legendre transform, that is, $h=e_\ell\circ\mathsf{leg}^{-1}$, we have
\[
h\left(\xi,\dot\xi,...,\xi^{(k-2)},\pi_{(1)},...,\pi_{(k-1)},\pi_{(0)}\right)=\sum_{j=0}^{k-1}\left\langle\pi_{(i)},\xi^{(i)}\right\rangle-\ell\left(\xi,\dot\xi,...,\xi^{(k-1)}\right)
\]
and one easily checks that these equations are equivalent to the $k^{th}$-order Euler-Poincar\'e equations \eqref{EP_k}. This follows from the definition of the momenta $\pi_{(i)}$, $i=1,..,k-1$, and the equalities
\begin{align*}
\frac{\delta h}{\delta\xi^{(i)}}&=\pi_{(i)}-\frac{\delta \ell}{\delta\xi^{(i)}},\quad i=0,...,k-2\\
\frac{\delta h}{\delta\pi_{(i)}}&=\xi^{(i)},\quad i=0,...,k-1.
\end{align*}

\paragraph{Example: Riemannian 2-splines on Lie groups.} The Legendre transform $\mathsf{leg}:3\mathfrak{g}\rightarrow T^*\mathfrak{g}\times \mathfrak{g}^*$ associated to the second order reduced Lagrangian 
\eqref{reduced-lag-ell} is 
\[
\mathsf{leg}\left(\xi,\dot\xi,\ddot\xi\right)=\left(\xi,\pi_{(1)},\pi_{(0)}\right),
\]
where
\[
\pi_{(1)}=\frac{\delta \ell}{\delta\dot\xi}=\dot\xi^\flat\pm\operatorname{ad}^*_\xi\xi^\flat=:\eta^\flat\quad \text{and}\quad\pi_{(0)}=\frac{\delta \ell}{\delta\xi}-\frac{d}{dt}\frac{\delta \ell}{\delta\dot\xi}=\mp\left(\operatorname{ad}^*_\eta\xi^\flat+(\operatorname{ad}_\eta\xi)^\flat\right)-\dot\eta^\flat.
\]
So the Hamiltonian formulation of the $2$-splines equations reads
\begin{equation}
\left\{\begin{array}{l}
\displaystyle \vspace{0.2cm}\left(\partial _t \pm \operatorname{ad}^*_{\frac{\delta  h}{\delta \pi _{(0)}} }\right)\pi _{(0)} =0
\,,\\
\displaystyle\partial _t \xi = \frac{\delta  h}{\delta \pi _{(1)}  },\qquad 
\partial _t \pi _{(1)}  = -\frac{\delta  h}{\delta \xi},
\end{array}\right.
\end{equation}
where the Hamiltonian $h:T^*\mathfrak{g}\times\mathfrak{g}^*$ is given by
\begin{align*}
h(\xi,\pi_{(1)},\pi_{(0)})&=\left\langle\pi_{(0)},\xi\right\rangle+\left\langle\pi_{(1)},\dot\xi\right\rangle-\ell(\xi,\dot\xi)\\
&=\frac{1}{2}\|\pi_{(1)}\|^2+\left\langle\pi_{(0)},\xi\right\rangle\mp \left\langle\pi_{(1)},(\operatorname{ad}^*_\xi\xi^\flat)^\sharp\right\rangle.
\end{align*}

\section{Ostrogradsky-Hamilton-Poincar\'e reduction}

In this Section we present the Hamiltonian side of the higher order Lagrange-Poincar\'e reduction carried out in Section \ref{Lagrange_Poincare_section}. It consists of a generalization of the Hamilton-Poincar\'e formulation to higher order, that uses Ostrogradsky momenta defined in \S\ref{Ostrogradsky_section} and encoded in the Legendre transform $\mathsf{Leg}:T^{(2k-1)}Q\rightarrow T^*\left(T^{(k-1)}Q\right)$. We refer to \cite{CeMaPeRa2003}, \S7, for the first order case, that is, the Hamilton-Poincar\'e reduction. As in \S\ref{Lagrange_Poincare_section}, we consider a free and proper action of a Lie group $G$ on the manifold $Q$ and we suppose that the Lagrangian $L:T^{(k)}Q\rightarrow\mathbb{R}$ is $G$-invariant under the action $\Phi^{(k)}$. In this case, the Poincar\'e-Cartan forms $\Theta_L$, $\Omega_L$, and the energy $E_L$ are $G$-invariant, that is,
\[
\left(\Phi^{(2k-1)}\right)^*\Theta_L=\Theta_L,\quad \left(\Phi^{(2k-1)}\right)^*\Omega_L=\Omega_L,\quad E_L\circ\Phi^{(2k-1)}=E_L
\]
and the Legendre transform is $G$-equivariant:
\[
\mathsf{Leg}\circ\Phi^{(2k-1)}=\left(\Phi^{(k-1)}\right)^{T^*}\circ\mathsf{Leg},
\]
see \cite{dLPiRo1994}, where $\left(\Phi^{(k-1)}\right)^{T^*}$ denotes the cotangent-lifted action of $\Phi^{(k-1)}$ on $T^*\left(T^{(k-1)}Q\right)$. Therefore, in the hyperregular case, the associated Hamiltonian $H$ is $G$-invariant: $H\circ \left(\Phi^{(k-1)}\right)^{T^*}=H$.

\paragraph{Legendre transform.} By fixing a principal connection $\mathcal{A}$ on the principal bundle $\pi:Q\rightarrow Q/G$, the reduced space on the Hamiltonian side can be identified with the bundle
\[
T^*\left(T^{(k-1)}(Q/G)\right)\oplus (k-1)\left(\operatorname{Ad}Q\oplus\operatorname{Ad}^*Q \right)\oplus \operatorname{Ad}^*Q
\]
over $\left(T^{(k-1)}Q\right)/G\simeq T^{(k-1)}(Q/G)\oplus (k-1)\operatorname{Ad}Q$. We will denote the reduced variables as
\[
\left(\rho,...,\rho^{(k-1)},\gamma_{(0)},...,\gamma_{(k-1)},\sigma,...,\sigma^{(k-2)},\pi_{(1)},...,\pi_{(k-1)},\pi_{(0)}\right)\mapsto \left(\rho,...,\rho^{(k-1)},\sigma,...,\sigma^{(k-2)}\right).
\]

Given a reduced Lagrangian $\ell:T^{(k)}(Q/G)\oplus k\operatorname{Ad}Q\rightarrow\mathbb{R}$, the reduced Legendre transform is the bundle map
\[
\mathsf{leg}:T^{(2k-1)}(Q/G)\oplus (2k-1)\operatorname{Ad}Q\rightarrow T^*\left(T^{(k-1)}(Q/G)\right)\oplus (k-1)\left(\operatorname{Ad}Q\oplus\operatorname{Ad}^*Q \right)\oplus \operatorname{Ad}^*Q
\]
covering the identity on $T^{(k-1)}(Q/G)\oplus (k-1)\operatorname{Ad}Q$ and given by
\[
\mathsf{leg}\left(\rho,...,\rho^{(2k-1)},\sigma,...,\sigma^{(2k-2)}\right)=\left(\rho,...,\rho^{(k-1)},\gamma_{(0)},...,\gamma_{(k-1)},\sigma,...,\sigma^{(k-2)},\pi_{(1)},...,\pi_{(k-1)},\pi_{(0)}\right),
\]
with
\begin{align}
\gamma_{(i)}&=\sum_{j=0}^{k-i-1}(-1)^j\frac{\partial^j}{\partial t^j}\frac{\partial \ell}{\partial \rho^{(i+j+1)}},\quad i=0,...,k-1\label{ostro_momenta1}\\
\pi_{(i)}&=\sum_{j=0}^{k-i-1}(-1)^j\frac{D^j}{D t^j}\frac{\delta \ell}{\delta \sigma^{(i+j)}},\quad i=0,...,k-1\label{ostro_momenta2}.
\end{align}
The reduced energy $e_\ell:T^{(2k-1)}(Q/G)\oplus (2k-1)\operatorname{Ad}Q\rightarrow\mathbb{R}$ is thus given by
\[
e_\ell\left(\rho,...,\rho^{(2k-1)},\sigma,...,\sigma^{(2k-2)}\right)=\sum_{i=0}^{k-1}\left\langle \gamma_{(i)},\rho^{(i+1)}\right\rangle+\sum_{i=0}^{k-1}\left\langle\pi_{(i)},\sigma^{(i)}\right\rangle-\ell\left(\rho,...,\rho^{(k)},\sigma,...,\sigma^{(k-1)}\right),
\]
where $\gamma_{(i)}$ and $\pi_{(i)}$ are given by \eqref{ostro_momenta1} and \eqref{ostro_momenta2}.

\paragraph{Reduced Hamilton equations.} In the hyperregular case, one obtains the reduced Hamiltonian by Legendre transform as $h=e_\ell\circ\mathsf{leg}^{-1}$ and we have
\begin{align}\label{relations_delta_l_h}
\frac{\partial h}{\partial\rho}&=-\frac{\partial\ell}{\delta\rho}\nonumber\\
\frac{\partial h}{\partial\rho^{(i)}}&=\gamma_{(i-1)}-\frac{\partial\ell}{\delta\rho^{(i)}},\quad i=1,...,k-1\nonumber\\
\frac{\partial h}{\partial\gamma_{(i)}}&=\rho^{(i+1)},\quad i=0,...,k-1\\
\frac{\delta h}{\delta\sigma^{(i)}}&=\pi_{(i)}-\frac{\delta \ell}{\delta\sigma^{(i)}},\quad i=0,...,k-2\nonumber\\
\frac{\delta h}{\delta\pi_{(i)}}&=\sigma^{(i)},\quad i=0,...,k-1.\nonumber
\end{align}
Using these relations and \eqref{ostro_momenta1}-\eqref{relations_delta_l_h}, we can rewrite the second equation of \eqref{Lagrange_poincare_eq} as
\begin{equation}\label{first_part}
\left(\frac{D}{Dt}\pm\operatorname{ad}^*_{\frac{\delta h}{\delta\pi_{(0)}}}\right)\pi_{(0)}=0,\;\; \frac{D}{Dt}\sigma^{(j-1)}=\frac{\delta h}{\delta\pi_{(j)}},\;\; \frac{D}{Dt}\pi_{(j)}=-\frac{\delta h}{\delta\sigma^{(j-1)}},\;\; j=1,...,k-1.
\end{equation}
With the help of \eqref{ostro_momenta2} and \eqref{relations_delta_l_h} we can write the right hand side of the first equation in \eqref{Lagrange_poincare_eq} as
\begin{align*}
&\sum_{j=0}^{k-1} \left\langle (-1)^j \frac{D^j}{Dt^j}\frac{\delta \ell }{\delta  \sigma ^ { (j) }}  \mp \sum_{p=0}^{j-1} (-1)^p \operatorname{ad}^*_{ \sigma ^{(j-1-p)} }\frac{D^p}{Dt^p} \frac{\delta \ell }{\delta  \sigma ^ { (j) }} ,\mathbf{i} _{\dot \rho}\tilde{\mathcal{B}}\right\rangle\\
&\displaystyle\quad =\left\langle\pi_{(0)}\mp \sum_{j=0}^{k-1}\sum_{p=0}^{j-1} (-1)^p \operatorname{ad}^*_{ \sigma ^{(j-1-p)} }\frac{D^p}{Dt^p}\left(\pi_{(i)}- \frac{\delta h }{\delta  \sigma ^ { (j) }} \right),\mathbf{i} _{\frac{\delta h}{\delta\gamma_{(0)}}}\tilde{\mathcal{B}}\right\rangle\\
&\quad =\left\langle\pi_{(0)}\mp \sum_{j=1}^{k-1}\operatorname{ad}^*_{ \sigma ^{(j-1)}}\pi_{(j)},\mathbf{i} _{\frac{\delta h}{\delta\gamma_{(0)}}}\tilde{\mathcal{B}}\right\rangle,
\end{align*}
where in the second line we used the convention that $\frac{\delta h }{\delta  \sigma ^ { (j) }}=0$ when $j=k-1$, and in the last equality we used the fact that many terms cancel out by the equation $\frac{D}{Dt}\pi_{(j)}=-\frac{\delta h}{\delta\sigma^{(j-1)}}$ in \eqref{first_part}.

Thus, we have shown that the $k^{th}$-order Lagrange-Poincar\'e equations \eqref{Lagrange_poincare_eq} can be rewritten in terms of $h$ as
\begin{framed}
\[
\left\{
\begin{array}{l}
\vspace{0.2cm}\displaystyle\frac{d}{dt}\gamma_{(0)}=-\frac{\delta h}{\delta \rho}-\left\langle \pi_{(0)}\mp \sum_{j=1}^{k-1}\operatorname{ad}^*_{ \sigma ^{(j-1)}}\pi_{(j)},\mathbf{i} _{\frac{\delta h}{\delta\gamma_{(0)}}}\tilde{\mathcal{B}}\right\rangle,\quad \frac{d}{dt}\rho=\frac{\delta h}{\delta \gamma_{(0)}}\\
\vspace{0.2cm}\displaystyle\frac{d}{dt}\gamma_{(i)}=-\frac{\delta h}{\delta\rho^{(i)}},\quad\frac{d}{dt}\rho^{(i)}=\frac{\delta h}{\delta\gamma_{(i)}},\quad i=1,...,k-1\\
\vspace{0.2cm}\displaystyle\left(\frac{D}{Dt}\pm\operatorname{ad}^*_{\frac{\delta h}{\delta\pi_{(0)}}}\right)\pi_{(0)}=0\\
\displaystyle\frac{D}{Dt}\sigma^{(i-1)}=\frac{\delta h}{\delta\pi_{(i)}},\quad\frac{D}{Dt}\pi_{(i)}=-\frac{\delta h}{\delta\sigma^{(i-1)}},\quad i=1,...,k-1.
\end{array}
\right.
\]
\end{framed}
We refer to this system as the {\bfi Ostrogradsky-Hamilton-Poincar\'e equations}. They can be obtained by Poisson reduction of the canonical Hamilton equations on $T^*(T^{(k-1)}Q)$, independently on a preexisting Lagrangian formulation. They reduce to the higher order canonical Hamilton equations when $G=\{e\}$, to the Hamilton-Poincar\'e equations when $k=1$, and to the Ostrogradsky-Lie-Poisson equations when $Q=G$.

For example, the Hamiltonian formulation of the second order Wong equations \eqref{2nd_order_Wong} reads
\[
\left\{
\begin{array}{l}
\vspace{0.2cm}\displaystyle\frac{d}{dt}\gamma_{(0)}=-\frac{\delta h}{\delta \rho}-\left\langle \pi_{(0)}\mp \operatorname{ad}^*_{\sigma}\pi_{(1)},\mathbf{i} _{\frac{\delta h}{\delta\gamma_{(0)}}}\tilde{\mathcal{B}}\right\rangle,\quad \frac{d}{dt}\rho=\frac{\delta h}{\delta \gamma_{(0)}}\\
\vspace{0.2cm}\displaystyle\frac{d}{dt}\gamma_{(1)}=-\frac{\delta h}{\delta\dot\rho},\quad\frac{d}{dt}\dot\rho=\frac{\delta h}{\delta\gamma_{(1)}}\\
\vspace{0.2cm}\displaystyle\left(\frac{D}{Dt}\pm\operatorname{ad}^*_{\frac{\delta h}{\delta\pi_{(0)}}}\right)\pi_{(0)}=0\\
\displaystyle\frac{D}{Dt}\sigma=\frac{\delta h}{\delta\pi_{(1)}},\quad\frac{D}{Dt}\pi_{(1)}=-\frac{\delta h}{\delta\sigma}.
\end{array}
\right.
\]

\paragraph{Poisson structure.} We now derive the reduced Poisson bracket obtained by Poisson reduction of the canonical symplectic form on $T^*(T^{(k-1)}Q)$. Given a function
\[
f=f\left(\rho,...,\rho^{(k-1)},\gamma_{(0)},...,\gamma_{(k-1)},\sigma,...,\sigma^{(k-2)},\pi_{(1)},...,\pi_{(k-1)},\pi_{(0)}\right)
\]
on the reduced space $T^*(T^{(k-1)}(Q/G))\oplus (k-1)\left(\operatorname{Ad}Q\oplus\operatorname{Ad}^*Q \right)\oplus \operatorname{Ad}^*Q$, its time derivative reads
\[
\dot f=\sum_{i=0}^{k-1}\left(\frac{\partial f}{\partial\rho^{(i)}}\dot\rho^{(i)}+\frac{\partial f}{\partial\gamma_{(i)}}\dot\gamma_{(i)}\right)+\sum_{j=1}^{k-1}\left(\frac{\delta f}{\delta\sigma^{(j-1)}}\frac{D}{Dt}\sigma^{(i-1)}+\frac{\delta f}{\delta\pi_{(j)}}\frac{D}{Dt}\pi_{(j)}\right)+\frac{\delta f}{\delta\pi_{(0)}}\frac{D}{Dt}\pi_{(0)}.
\]
Using this formula and the Ostrogradsky-Hamilton-Poincar\'e equations, one obtains the following expression for the Poisson bracket
\begin{align*}
\{f,g\}&=\{f,g\}_{can}\left(\rho,...,\rho^{(k-1)},\gamma_{(0)},...,\gamma_{(k-1)}\right)+\{f,g\}_{can}\left(\sigma,...,\sigma^{(k-2)},\pi_{(1)},...,\pi_{(k-1)}\right)\\
&\qquad\pm\left\langle\pi_{(0)},\left[\frac{\delta f}{\delta\pi_{(0)}},\frac{\delta g}{\delta\pi_{(0)}}\right]\right\rangle+\left\langle\pi_{(0)}+\sum_{j=1}^{k-1}\operatorname{ad}^*_{\sigma^{(j-1)}}\pi_{(j)},\tilde{\mathcal{B}}\left(\frac{\partial f}{\partial\gamma_{(0)}},\frac{\partial g}{\partial\gamma_{(0)}}\right)\right\rangle.
\end{align*}
In the first order case, this bracket consistently recovers the \textit{gauged-Lie-Poisson} bracket introduced in \cite{MoMaRa1984}; see also \cite{Mo1986} and \cite{CeMaPeRa2003}.

\section{Conclusions}

This paper has developed the higher-order framework for Lagrangian and Hamiltonian reduction by symmetry in geometric mechanics. The extension of the framework of geometric mechanics is made to the higher order case when the Lagrangian function is defined on the $k^{th}$-order tangent bundle $T^{(k)}Q$ and thus depends on the first $k^{th}$-order time derivatives of the curve.  The \textit{$k^{th}$-order Lagrange-Poincar\'e equations} on $T^{(k)}Q/G$ have been derived and the \textit{$k^{th}$-order Euler-Poincar\'e equations} on $T^{(k)}G/G\simeq k\mathfrak{g}$ have been obtained in the particular case $Q=G$, together with the associated constrained variational formulations.

On the Hamiltonian side, the Legendre transform $T^{(2k-1)}Q\rightarrow T^*\left(T^{(k-1)}Q\right)$ associated to the Ostrogradsky momenta has been applied to obtain what we call the \textit{Ostrogradsky-Hamilton-Poincar\'e equations} on $T^*\left(T^{(k-1)}Q\right)/G$ and, in the particular case $Q=G$, the \textit{Ostrogradsky-Lie-Poisson equations}.

These are only the first steps needed in developing the higher-order framework of geometric mechanics and reduction by symmetry for applications, particularly in the registration of a sequence of images, as begun in \cite{GBHoMeRaVi2011}. 

In this paper we only considered unconstrained dynamics. It would be interesting to generalize the ideas developed here to the case of  higher order nonholonomic systems and vakonomic systems, by extending the approach of  \cite{CeGr2007}, 
\cite{Grillo2009}, and \cite{CoMa2011}.

{\footnotesize

\bibliographystyle{new}
 
}

\end{document}